\documentclass{iopart}

\usepackage{epsf}

\newcommand\be{\begin{equation}}
\newcommand\ee{\end{equation}}

\begin{document}


\title{Identified particles from viscous hydrodynamics}

\author{Denes Molnar}

\address{Physics Department, Purdue University, 
West Lafayette, IN 47907-2036, U.S.A.}
\begin{abstract}
Identified particle observables from viscous hydrodynamics are sensitive
to the fluid-to-particle conversion.
Instead of the commonly assumed ``democratic'' 
Grad ansatz for phase space corrections $\delta f$, we utilize corrections
calculated from linearized covariant
transport theory. Estimates based on a $\pi-p$ system with binary collisions
indicate
that protons are much closer to equilibrium than pions, significantly
affecting the dissipative reduction of differential elliptic flow 
in Au+Au at RHIC. In addition, we test linear response against fully nonlinear
transport for a two-component massless system in a Bjorken scenario.
Strikingly, we find that, while linear response accounts well for the dynamical
sharing of shear stress,
the momentum dependence of phase space corrections is best described by
Grad's quadratic ansatz, and not the linear response solution.
\end{abstract}

%
%

\section{Introduction}

An inevitable ingredient in contrasting hydrodynamic calculations of 
heavy-ion collisions to experimental data is the conversion of the fluid to 
particles (hadrons).
For ideal fluids this is straightforward, at least within
the Cooper-Frye framework\cite{CooperFrye}.
For viscous fluids, however, dissipative corrections such as shear stress,
distort phase space distributions
from local equilibrium,
$f = f_{eq} + \delta f$,
and the very 
same viscous hydrodynamic fields can be described by an infinite class of
phase space corrections.

Viscous hydrodynamic 
calculations\cite{LuzumRomatschke,HeinzSong,Schenke} commonly 
ignore this ambiguity and, in the presence of shear stress,
pick corrections
for each hadron species in the ``democratic'' form (\ref{df_democ}).
This ad-hoc choice does not
reflect that species that interact more frequently 
should be closer to equilibrium.

We use here covariant transport theory to 
compute dissipative phasespace corrections in a hadron gas.
In the linear response limit, quite akin
to \cite{TeaneyQG},
we obtain first results for a pion-proton system with binary $2\to 2$ rates, 
illustrating how 
dynamical sharing of shear stress between species affects
identified particle differential elliptic flow $v_2(p_T)$.
We also test the validity of linear response against 
fully {\em nonlinear} $2\to 2$ transport for 
a two-component mixture undergoing a 0+1D 
Bjorken
expansion.

\section{Dissipative phase space corrections and covariant transport}

For zero bulk viscosity,
the only dissipative field in viscous hydrodynamics
is shear stress $\pi^{\mu\nu}(x)$ and therefore
dissipative corrections for species $i$ are of the form
$\delta f_i = C_i(p_{i\alpha} u^\alpha / T)  \pi^{\mu\nu}p_{i\mu} p_{i\nu} f_{i,eq}$, where $u$ is the flow velocity (Boltzmann statistics used). 
$C_i$ are arbitrary
functions with the only constraint
that partial shear stresses sum to the total
$\sum\limits \pi^{\mu\nu}_i = \pi^{\mu\nu}$, naturally 
satisfied by the ``democratic'' Grad prescription
\be
C_i^{dem.} \equiv \frac{1}{2(e+p)T^2}  \qquad \Rightarrow \qquad 
\delta f_i^{dem.} = \frac{\pi^{\mu\nu} 
p_{i\mu} p_{i\nu}}{2T^2 (\epsilon+p)} f_{i,eq}
\label{df_democ}
\ee

On the other hand, from covariant transport 
one expects $\delta f_i$ to depend on the mean free paths,
or equivalently
the inverse Knudsen numbers $K_i \sim \sum\limits_j n_j 
\langle\sigma v_{rel}\rangle_{ij} / \tau$.
We consider on-shell\cite{visc} $2\to 2$ Boltzmann 
transport for each species.

Linear response to shear is the same expansion around local equilibrium
as the standard calculation of shear 
viscosity\cite{DeGroot,AMY}, leading to an integral equation 
for $\{\delta f_i\}$ that is solved variationally,
resulting in corrections proportional to spatial gradients in the comoving
frame (as in Navier-Stokes hydrodynamics).
Because {\em nonequilibrium terms are only kept in the collision
kernel}, this ignores shear stress relaxation
captured by second-order
hydrodynamics\cite{IS,isvstr,LuzumRomatschke,HeinzSong,TeaneyQG}.

Alternatively, we solve the fully nonlinear transport
for a two-component mixture
that evolves near equilibrium. This way we study $\delta f$ in a less
constrained dynamical setting, which also provides useful cross-checks 
for the
linear response results.

\begin{figure}[h]
\leavevmode
\begin{center}
\epsfysize=5.6cm
\epsfbox{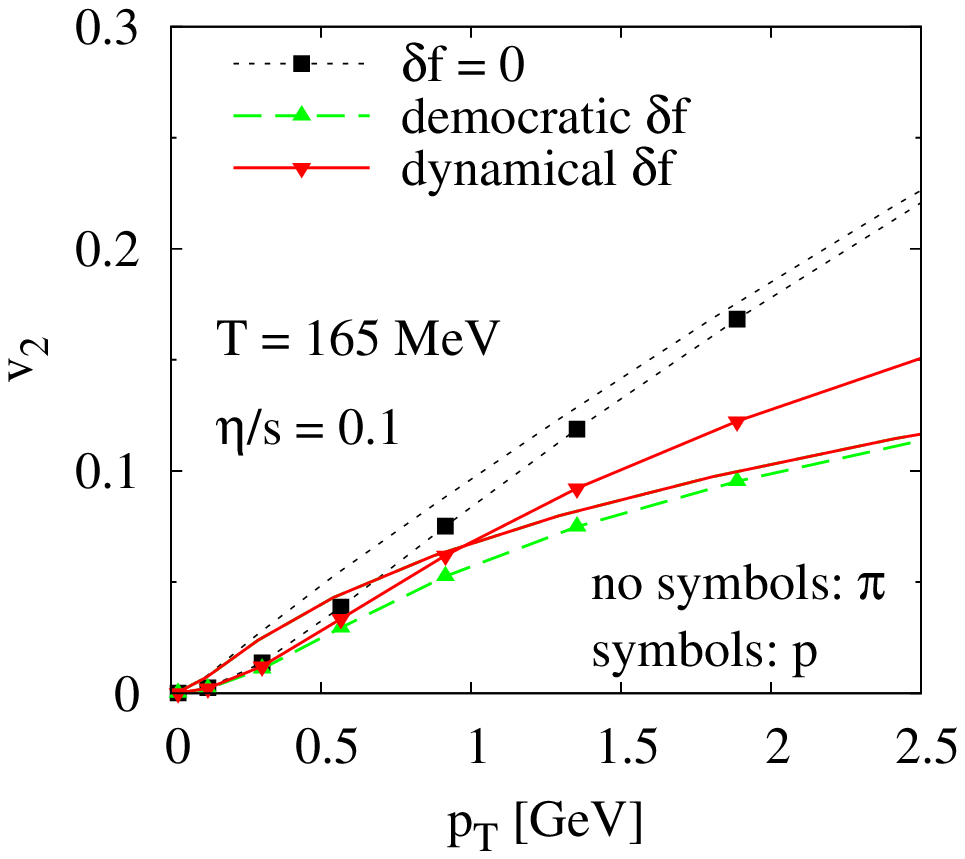}
\hskip 0cm
\epsfysize=5.6cm
\epsfbox{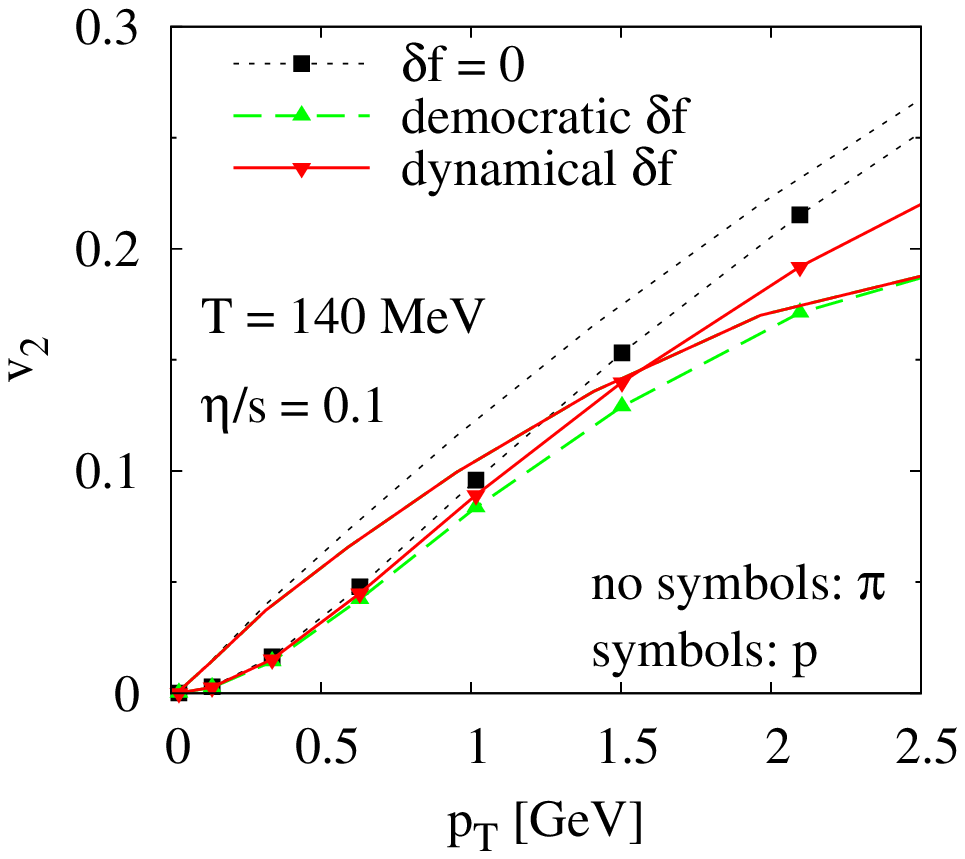}
\caption{Differential elliptic flow $v_2(p_T)$ of protons (symbols) 
and pions (lines with no symbols) 
for various dissipative corrections $\delta f$ and
freezeout temperatures $T_{fo} =  165$ MeV (left) and 140 MeV (right). 
Dotted: ideal fluid ($\delta f = 0$), dashed: 
viscous fluid with ``democratic'' $\delta f$, 
solid: viscous fluid with $\delta f$ determined dynamically for a $\pi-p$ 
mixture. In both viscous cases, we take the Grad ansatz 
$\delta f \propto p^2$ and set the same partial shear stress for pions
(i.e., the solid and dashed lines without symbols coincide).
The $\pi-p$ splitting pattern is sensitive to the 
$\pi$ and $p$ equilibration rates, especially at higher $T_{fo}$.
}
\label{fig:1}
\end{center}
\vspace*{-0.3cm}
\end{figure}

\section{Highlighted results}

Figure~\ref{fig:1} shows how dynamical shear stress sharing influences 
differential elliptic flow $v_2(p_T)$ at freezeout in Au+Au at RHIC.
We focus here on a $\pi-p$ system, i.e., ignore
interactions of protons and pions with other hadrons, 
and also ignore resonance decays.
The ``dynamical'' results are from linear response with
isotropic, energy-independent effective scattering cross sections 
$\sigma_{\pi\pi} = 30$~mb, $\sigma_{\pi p} = 20$~mb, $\sigma_{p p} = 50$~mb.
At $T \sim 120-165$~MeV, 
these approximate quite well a more realistic
calculation\cite{Prakash_pip} of mean collision
frequencies in a $\pi-p$ system
based on measured phase shifts.
For simplicity, 
quadratic momentum dependence was assumed for both species,
however, $C_\pi$ and $C_p$ were optimized variationally, i.e., $C_\pi \ne C_p$.
The ``democratic'' prescription corresponds to quadratic momentum
dependence with $C_\pi = C_p$.
Results for ideal hydrodynamics ($\delta f = 0$) are also shown.
Ideal hydrodynamic fields for Au+Au at $\sqrt{s}=200A$ GeV, $b = 7$ fm,
were obtained with our patched version {\sf 0.2p2} 
of the 2+1D boost-invariant {\sf AZHYDRO} code\cite{AZHYDRO}.
We used the recent {\sf s95-p1}
equation of state parameterization\cite{EOSs95p1}
matching lattice QCD results to a hadron resonance gas.
Shear stress at freezeout was estimated as in\cite{Derek_olddeltaf},
based on gradient corrections to ideal hydro
$\pi^{\mu\nu}
= \eta [\nabla^\mu u^\nu
             + \nabla^\nu u^\mu
             - (2/3)\Delta^{\mu\nu} (\partial_\alpha u^\alpha)
      ]$ with $\Delta^{\mu\nu} \equiv g^{\mu\nu} - u^\mu u^\nu$, 
$\nabla^\mu \equiv \Delta^{\mu\nu} u_\nu$. But unlike\cite{Derek_olddeltaf},
we use real hydro instead of the parameterizations (``Blast-wave'').

The left plot in Figure~\ref{fig:1} shows the $\pi-p$ splitting of $v_2(p_T)$
for 
freezeout at $T = 165$ MeV, the typical switching
temperature between hydrodynamics and transport in hydro+transport 
calculations\cite{VISHNU}.
At moderate 
$p_T \sim 1-2$~GeV,
we find a significant viscous reduction of elliptic flow
for both species even for a small shear viscosity to entropy density ratio 
$\eta/s = 0.1$.
However, compared with the dynamical approach
in which protons have smaller viscous correction than pions,
the ``democratic'' prescription 
oversuppresses proton elliptic flow by $20-30$\%.
In the dynamical approach, pion-proton crossing occurs at much lower 
$p_T \sim 1$ GeV.
At lower $T = 140$ MeV (right plot), 
more applicable to pure hydrodynamic description
of RHIC data, dissipative corrections are smaller,
but the relative difference between the ``dynamical'' and
``democratic'' approaches is in fact larger.

\begin{figure}
\leavevmode
\begin{center}
\epsfysize=5.6cm
\epsfbox{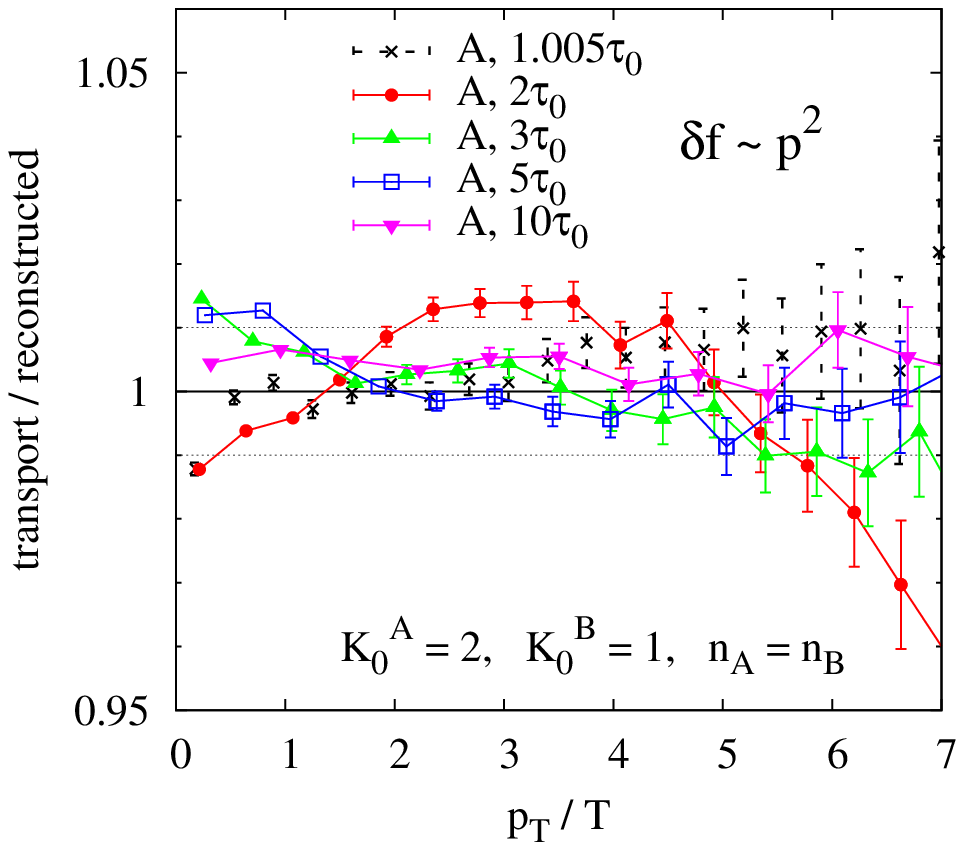}
\epsfysize=5.6cm
\epsfbox{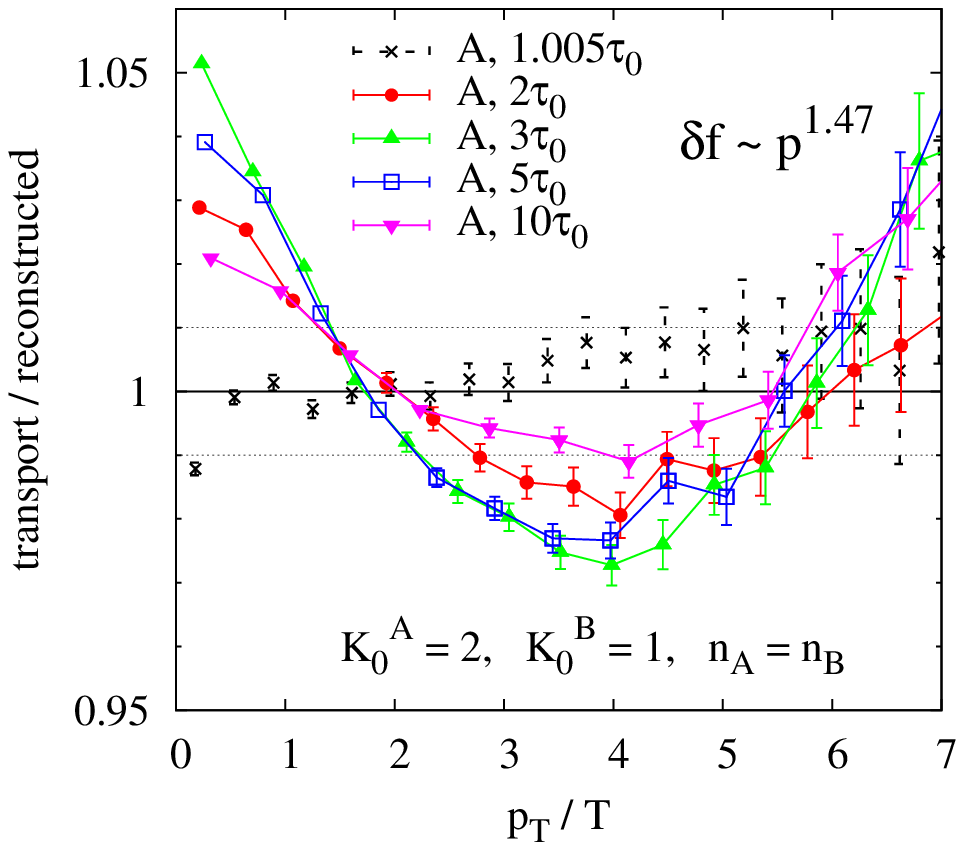}
\caption{Ratio of spectra from covariant transport to reconstructed spectra
using only the partial shear stresses from the transport 
and an {\em ansatz} for $\delta f$,
for a two-component massless system in a 0+1D Bjorken scenario.
The ratio is plotted at various proper times 
$\tau = const \times \tau_0$ versus normalized transverse momentum 
$p_T /T(\tau)$, for species $A$, which is set to scatter twice as often as $B$ (the ratio of inverse Knudsen numbers is $K_A/K_B = 2$).
We find that Grad's quadratic ansatz for the momentum dependence (left plot)
is much more accurate than 
the momentum dependence computed from linear response (right plot).
}
\label{fig:2}
\end{center}
\vspace*{-0.3cm}
\end{figure}

Figure~\ref{fig:2} shows dissipative corrections from fully nonlinear $2\to 2$
transport for a two-component
massless gas $A+B$ 
undergoing 0+1D Bjorken expansion (transverse translational symmetry assumed). 
$f_A$ and $f_B$ only depend on normalized proper time 
$\tau/\tau_0$, 
$p_T$, and the difference between coordinate and momentum rapidity.
We set isotropic 
$\sigma_{AA} : \sigma_{AB} : \sigma_{BB} = 4 : 2 : 1$, densities $n_A = n_B$,
such that (initial) inverse Knudsen numbers are $K_A(\tau_0) = 2$, 
$K_B(\tau_0) = 1$, i.e., species $A$ scatters twice as often as $B$,
and start from local equilibrium.
To keep $\eta/s \approx const$,
we scale\cite{visc} cross sections with time as $\sigma \propto \tau^{2/3}$.
We test how well various {\em ans\"atze} 
for $\delta f$ can reconstruct, {\em solely} from the temperature and
partial shear stresses, the spectra from the transport.

With the dynamically determined shear stresses from the transport,
Grad's quadratic ansatz $\delta f \propto p^2$ (left plot)
is accurate to within about a 
percent(!) for a wide range of times (up to $10\tau_0$), and momenta 
(up to $5-6\times T$). However, with the
momentum dependence $\delta f \propto p^{1.47}$ 
calculated from linear response (right plot), the agreement is much poorer.
The likely reason is that shear stress relaxation competes with the
rapid expansion of the system, and therefore, relaxation to Navier-Stokes
should not be assumed. Note, the power $1.47$ we obtained for
isotropic $2\to 2$ processes is quite similar to the one in \cite{TeaneyQG} 
for forward-peaked $2\to 2$ and radiative $1\leftrightarrow 2$.

\section{Conclusions}

Reliable extraction of shear viscosity from heavy-ion data must
involve proper conversion of viscous fluid to hadrons.
Though several aspects of this work need to be refined in the future
(e.g., with realistic cross sections, 
real viscous hydro, quantum statistics, more species),
our results do indicate that the widely used conversion prescription 
(``democratic'' Grad approach)
is inconsistent with scattering rates in a hadron gas.
This brings into question the accuracy of viscous hydrodynamic 
calculations of identified
particle observables to date.

\ack This work was supported by the US DOE under contract DE-PS02-09ER41665.



\section*{References}

\begin{thebibliography}{10}
\bibitem{CooperFrye}
Cooper~F and Frye~G 1974
{\it Phys. Rev.} D {\bf 10} 186.
For a discussion of some of the limitations,
see, e.g.,
  Grassi~F, Hama~Y and Kodama~T 1995
  {\it Phys.\ Lett.}\  B {\bf 355} 9;
  Molnar~D and Gyulassy~M 2000
  {\it Phys.\ Rev.}\ C {\bf 62} 054907 
  ({\it Preprint} arXiv:nucl-th/0005051); and
  Bugaev~K~A 2003
  {\it Phys.\ Rev.\ Lett.}\  {\bf 90} 252301
  ({\it Preprint} arXiv:nucl-th/0210087)


\bibitem{LuzumRomatschke}
  Luzum~M and Romatschke~P 2008
  {\it Phys.\ Rev.}\  C {\bf 78} 034915
  ({\it Erratum} 2009 {\it Phys.\ Rev.}\   C {\bf 79} 039903)
  ({\it Preprint} arXiv:0804.4015 [nucl-th])

\bibitem{HeinzSong}
  Shen~C, Heinz~U, Huovinen~P and Song~H 2010,
 {\it Phys.\ Rev.}\  C {\bf 82} 054904
  ({\it Preprint} arXiv:1010.1856 [nucl-th])

\bibitem{Schenke}
  Schenke~B, Jeon~S and Gale~C 2010,
  {\it Phys.\ Rev.}\  C {\bf 82} 014903
  ({\it Preprint} arXiv:1004.1408 [hep-ph])


\bibitem{TeaneyQG}
  Dusling~K, Moore~G~D and Teaney~D 2010,
  {\it Phys.\ Rev.}\ C {\bf 81} 034907
  ({\it Preprint} arXiv:0909.0754 [nucl-th])


\bibitem{visc}
  Molnar~D 2008
  {\it Preprint} arXiv:0806.0026 [nucl-th]

\bibitem{DeGroot}
de~Groot~S~R, van~Leeuwen~W~A, van~Weert~Ch~G 1980
{\it Relativistic kinetic theory - Principles and applications}
(North-Holland)

\bibitem{AMY}
  Arnold~P~B, Moore~G~D and Yaffe~L~G 2000
  {\it JHEP} {\bf 0011} 001
  ({\it Preprint} arXiv:hep-ph/0010177)


\bibitem{IS} 
Israel~W, Stuart~J~M 1979
{\it Ann.\ Phys.} {\bf 118} 349

\bibitem{isvstr}
  Huovinen~P and Molnar~D 2009
  {\it Phys.\ Rev.}\  C {\bf 79} 014906
  ({\it Preprint} arXiv:0808.0953 [nucl-th])



\bibitem{Prakash_pip}
  Prakash~M, Prakash~M, Venugopalan~R and Welke~G 1993
  {\it Phys.\ Rept.}\  {\bf 227} 321

\bibitem{AZHYDRO}
Kolb~P~F, Sollfrank~J, and Heinz~U 2000 {\it Phys.\ Rev.}\ C {\bf 62} 054909;
Kolb~P~F and Heinz~U 2003 {\it Preprint nucl-th/0305084}.
The original version 0.2 of the code and versions patched by
D.~Molnar and P.~Huovinen are available from the
OSCAR repository at http://karman.physics.purdue.edu/OSCAR

\bibitem{EOSs95p1}
 Huovinen~P and Petreczky~P 2010
  {\it Nucl.\ Phys.}\  A {\bf 837} 26
  ({\it Preprint} arXiv:0912.2541 [hep-ph])

\bibitem{Derek_olddeltaf}
  Teaney~D 2003
  {\it Phys.\ Rev.}\  C {\bf 68} 034913
  ({\it Preprint} arXiv:nucl-th/0301099)

\bibitem{VISHNU}
  Song~H, Bass~S~A and Heinz~U 2011
  {\it Phys.\ Rev.}\  C {\bf 83} 024912
  ({\it Preprint} arXiv:1012.0555 [nucl-th])


\end{thebibliography}
\end{document}